\documentclass[twocolumn]{aastex62}

\usepackage[autostyle]{csquotes}
\usepackage{cancel}
\shorttitle{Time \& Charge-Sign Dependence of Modulation}
\shortauthors{Aslam et al.}

\begin{document}

\title{Time and Charge-Sign Dependence of the Heliospheric Modulation of Cosmic Rays}
\email{aslamklr2003@gmail.com (OPMA) \\
}

\author[0000-0001-9521-3874]{O.P.M. Aslam}
\affil{Centre for Space Research, North-West University, 2520 Potchefstroom, South Africa \\}

\author[0000-0001-7623-9489]{D. Bisschoff}
\affil{Centre for Space Research, North-West University, 2520 Potchefstroom, South Africa \\}

\author[0000-0001-5844-3419]{M. D. Ngobeni}
\affil{Centre for Space Research, North-West University, 2520 Potchefstroom, South Africa \\}
\affil{School of Physical and Chemical Sciences, North-West University, 2735 Mmabatho, South Africa\\}

\author[0000-0003-0793-7333]{M. S. Potgieter}
\affil{Retired; FS4, 2531 Potchefstroom, South Africa \\}

\author[0000-0001-7598-1825]{R. Munini}
\affil{INFN, Sezione di Trieste I-34149 Trieste, Italy \\}

\author[0000-0002-8015-2981]{M. Boezio}
\affil{INFN, Sezione di Trieste I-34149 Trieste, Italy \\}

\author[0000-0003-3851-2901]{V. V. Mikhailov}
\affil{National Research Nuclear University, MEPHI, RU-115409, Moscow, Russia\\}

\begin{abstract}

Simultaneous and continuous observations of galactic cosmic-ray  electrons ($e^{-}$) and positrons ($e^{+}$) from the $\it{PAMELA}$ and $\it{AMS}02$ space experiments are most suitable for numerical modeling studies of the heliospheric modulation of these particles below 50 GeV. A well-established comprehensive three-dimensional (3D) modulation model is applied to compute full spectra for $e^{-}$ and $e^{+}$ with the purpose of reproducing the observed ratio $e^{+}$/$e^{-}$ for a period which covers the previous long and unusual deep solar minimum activity and the recent maximum activity phase including the polarity reversal of the solar magnetic field. For this purpose the very local interstellar spectra for these particles were established first. Our study is focused on how the main modulation processes, including particle drifts, and other parameters such as the three major diffusion coefficients, had evolved, and how the corresponding charge-sign dependent modulation had occurred subsequently. The end result of our effort is the detailed reproduction of $e^{+}$/$e^{-}$ from 2006 to 2015, displaying both qualitative and quantitative agreement with the main observed features. Particularly, we determine how much particle drifts is needed to explain the time dependence exhibited by the observed $e^{+}$/$e^{-}$ during each solar activity phase, especially during the polarity reversal phase when no well-defined magnetic polarity was found.
    
\end{abstract}

\keywords{Galactic cosmic rays; Particle astrophysics; Heliosphere; Solar activity}

\section{Introduction} \label{sec1}

The continuously changing behavior of the Sun and its activity cycle have kept the heliospheric modulation of galactic cosmic-rays (GCRs) a subject of interest and intense research. Recent GCR observations at the Earth reported by the $\it{PAMELA}$ (Adriani et al. 2017; Boezio et al. 2017) and $\it{AMS}02$ experiments (Aguilar et al. 2018a, b), together with $\it{Voyager}$ 1 (Stone et al. 2013; Cummings et al. 2016) and $\it{Voyager}$ 2 (Stone et al. 2019) observations in the outer heliosphere, have contributed significantly to major developments in solar modulation studies, especially in the numerical modeling of the modulation of these GCR particles. Measurements beyond the heliopause by $\it{Voyager}$ 1 and 2 have assisted substantially to determine and present very local interstellar spectra (VLIS) for GCRs with confidence down to a few MeV/nuc (Potgieter 2014a; Cummings et al. 2016; Boschini et al. 2017; Bisschoff et al. 2019).

GCRs are modulated by the turbulent solar wind and the embedded heliospheric magnetic field (HMF) when they enter the heliosphere, where the transport of these particles are subjected to the following physical process: (i) Convection caused by the outward directed solar wind velocity. (ii) Gradient, curvature and heliospheric current sheet (HCS) drifts. (iii) Adiabatic energy changes because of the expanding solar wind, depending on the sign of the divergence of the solar wind velocity. (iv) Spatial diffusion caused by scattering off random magnetic field irregularities. In this context, see reviews by Heber (2013), K\'{o}ta (2013) and Potgieter (1998, 2013, 2017). All these physical processes are considered important in modulating GCRs, although their relative importance and how this changes with solar activity and with reversals of the HMF polarity (e.g. Webber et al. 1990; Heber et al. 2002, 2003; McDonald et al. 2010; Aslam \& Badruddin 2014) is still intensively investigated.
Potgieter (2014b) concluded that the heliospheric modulation of GCRs is always an intriguing interplay of these four basic physical mechanisms that changes with solar activity and the polarity cycles. Based on this view point, it is expected that particle drifts may play a somewhat dominant role during low to moderate solar activity periods, dependent on the direction (polarity) of the HMF, so that a different modulation effect is caused and observed during consecutive solar polarity cycles, while diffusion may become more dominant during periods of maximum solar activity, not determined by the solar magnetic polarity (see also Pacini \& Usoskin 2015; Aslam et al. 2019a). Adiabatic energy loss is always important, at the Earth (deep inside the heliosphere) even dominating the shape of the spectra of GCR protons (and anti-protons), also light and heavier nuclei at lower kinetic energies, but far less in the case of GCR electrons ($e^{-}$) and positrons ($e^{+}$); see illustrations by Moraal \& Potgieter (1982).   

Around the maximum phase of solar activity, the HMF reverses its polarity. When these field lines are directed outward in the Sun's southern hemisphere and inward in the northern hemisphere, as occurred from 2001 to 2012, it is termed as a negative polarity phase (A$<$0), whereas when the field lines are directed outward in the Sun's northern hemisphere and inward in the southern hemisphere, it is termed a positive polarity phase (A$>$0), which has been the case since April 2014. The HMF polarity had reversed from A$<$0 to A$>$0 between November 2012 and March 2014 (see Sun et al. 2015; also, solar polar field observations, http://wso.stanford.edu). The period between these dates is considered as a phase of the HMF without a well-defined polarity. In an A$<$0 polarity phase (2001-2012) positively charged GCRs (protons, positrons, helium nuclei, etc.) drift inward to the inner heliosphere mainly through the equatorial regions, and in the process encounter the wavy HCS, and outward via the polar regions of the heliosphere, whereas negatively charged particles (electrons, anti-protons, all anti-matter nuclei) then drift downward toward the Earth mainly from the polar regions and then outward mainly through the equatorial regions. During this phase, it is expected that the changing HCS may play an important, even dominant, role in the modulation of positively charged particles, whereas for an A$>$0 polarity phase the drift directions become reversed so that negatively charged particles encounter the HCS during their entry (Jokipii \& Thomas 1981; K\'{o}ta \& Jokipii 1983; Potgieter \& Moraal 1985; Webber et al. 1990; Potgieter \& le Roux 1992; Aslam et al. 2019a, and references there-in). This produces a 22-year modulation cycle and causes a charge-sign dependent effect in the solar modulation of GCRs as reviewed by Potgieter (2014a).    
 
In the context of charge-sign dependent modulation, the modeling study presented here is specifically focused on the ratio $e^{+}$/$e^{-}$, observed by the $\it{PAMELA}$ experiment from July 2006 to December 2015 as reported by Adriani et al. (2016a) and similar observations after May 2011 by the $\it{AMS}02$ experiment (Aguilar et al. 2018b). In order to compute these ratios with numerical models for a Kinetic Energy (KE) range relevant to solar modulation, a detailed modulation study of the corresponding electron and positron spectra must be made for the mentioned time period. The modulation of $e^{+}$ and $e^{-}$ differ in the way that they experience drifts, and in the shape of their VLIS, of course.

For such an effort, we utilized the published 6-month averaged $\it{PAMELA}$ $e^{-}$ spectra (Adriani et al. 2015) and $e^{+}$ spectra (Munini 2015) for July 2006 to December 2009, and $e^{-}$ and $e^{+}$ spectra averaged over Bartel rotation periods observed by $\it{AMS}02$ from May 2011 to December 2015 (Aguilar et al. 2018b), along with our comprehensive three-dimensional (3D) numerical modulation model as will be described below. This particular model includes gradient, curvature and HCS drift, and have been applied successively earlier and also recently to the modulation studies of GCRs such as protons (Potgieter et al. 2014; Vos \& Potgieter 2015), electrons (Potgieter et al. 2015; Potgieter \& Vos 2017), positrons (Potgieter et al. 2017; Aslam et al. 2019a), and helium nuclei (Ngobeni et al. 2020). All these studies are related to observations by the $\it{PAMELA}$ experiment for the unusual solar minimum of 2006 to 2009 (Adriani et al. 2013a, 2015; Munini et al. 2015, 2017; Marcelli et al. 2020). Corti et al. (2019) applied the same numerical code to study the modulation of galactic protons and helium observed by the $\it{AMS}02$ (Aguilar et al. 2018a) for 2011 to 2017. 

The period July 2006 to December 2015 includes the unusual deep and long solar minimum of 2006 to 2009 and the subsequent solar activity maximum phase with the solar magnetic polarity reversal period of 2012 to 2014. These continuous and simultaneous observations of $e^{-}$ and $e^{+}$ fluxes below 50 GeV is most suitable for comprehensive solar modulation studies in the inner heliosphere and have served as a strong motivation to focus our modeling efforts on this relatively long period to obtain insight on how these spectra had evolved and how the corresponding charge-sign dependence had unfolded. This period, as long as a solar cycle, is particularly challenging from a modeling point of view because it includes a solar minimum modulation period, a period of maximum modulation, both the HMF polarity phases, the switch from A$<$0 to A$>$0, which was separated by a $\approx$ 1.5-year long period with no well-defined HMF polarity. 

 \section{Positron and Electron Observations from $\it{PAMELA}$ and $\it{AMS}02$} \label{sec2}

Since the beginning of the $\it{PAMELA}$ space experiment in June 2006, and $\it{AMS}$02 in May 2011, continuous and precision measurements of primary and secondary components of GCR flux of high statistics, over a wide energy range and with different time resolution have been reported. Picozza et al. (2007), Adriani et al. (2014, 2017), and Boezio et al. (2017) explained in detail the features of the $\it{PAMELA}$ experiment, and the 10-year achievements of the mission. Adriani et al. (2009a) reported the first results of $\it{PAMELA}$ on the $e^{+}$ fraction over an energy range of 1.5 - 100 GeV for the period July 2006 to February 2008. As a continuation, observations of anti-protons (Adriani et al. 2009b, 2010), protons and several isotopes were reported with a varying time resolution (Adriani et al. 2011a, 2013a, 2016b), also Carrington rotation and longer averaged helium (Adriani et al. 2011a; Marcelli et al. 2020), 6-months and longer averaged electrons (Adriani et al. 2011b, 2015) and positrons (Adriani et al. 2013b, 2016a; Munini, 2015).   

The first results of the $\it{AMS}$02 experiment, located on the International Space Station (ISS), of precision measurements of the $e^{+}$ fraction up to a few 100 GeV was reported by Aguilar et al. (2013). Recent publications from this mission reported GCR fluxes from May 2011 to May 2017, averaged over Bartel rotations, in particular for protons and helium (Aguilar et al. 2018a), electrons and positrons (Aguilar et al. 2018b), all above 1.0 GeV.     
 
Of specific interest for this modeling work is the $\it{PAMELA}$ observed $e^{+}/e^{-}$ from July 2006 to December 2015 for energies between 0.5 GeV and 5.0 GeV as reported by Adriani et al. (2016a). It was the first efficient and continuous observation of this ratio over a period almost as long as a solar cycle, including the quiet and unusual solar minimum of solar cycles 23 to 24, followed by solar maximum activity and the HMF polarity reversal phase of solar cycle 24, as mentioned above. However, full spectra for these particles over the complete operational time of this mission have not yet been published. $\it{AMS}02$ observations of $e^{-}$ and $e^{+}$ fluxes, and the ratio $e^{+}/e^{-}$, were reported by Aguilar et al. (2018b) from May 2011 to May 2017 (Bartel rotation averaged: from Bartel rotation number 2426 to 2506).    
      
\begin{figure*}[!htp]
\plotone{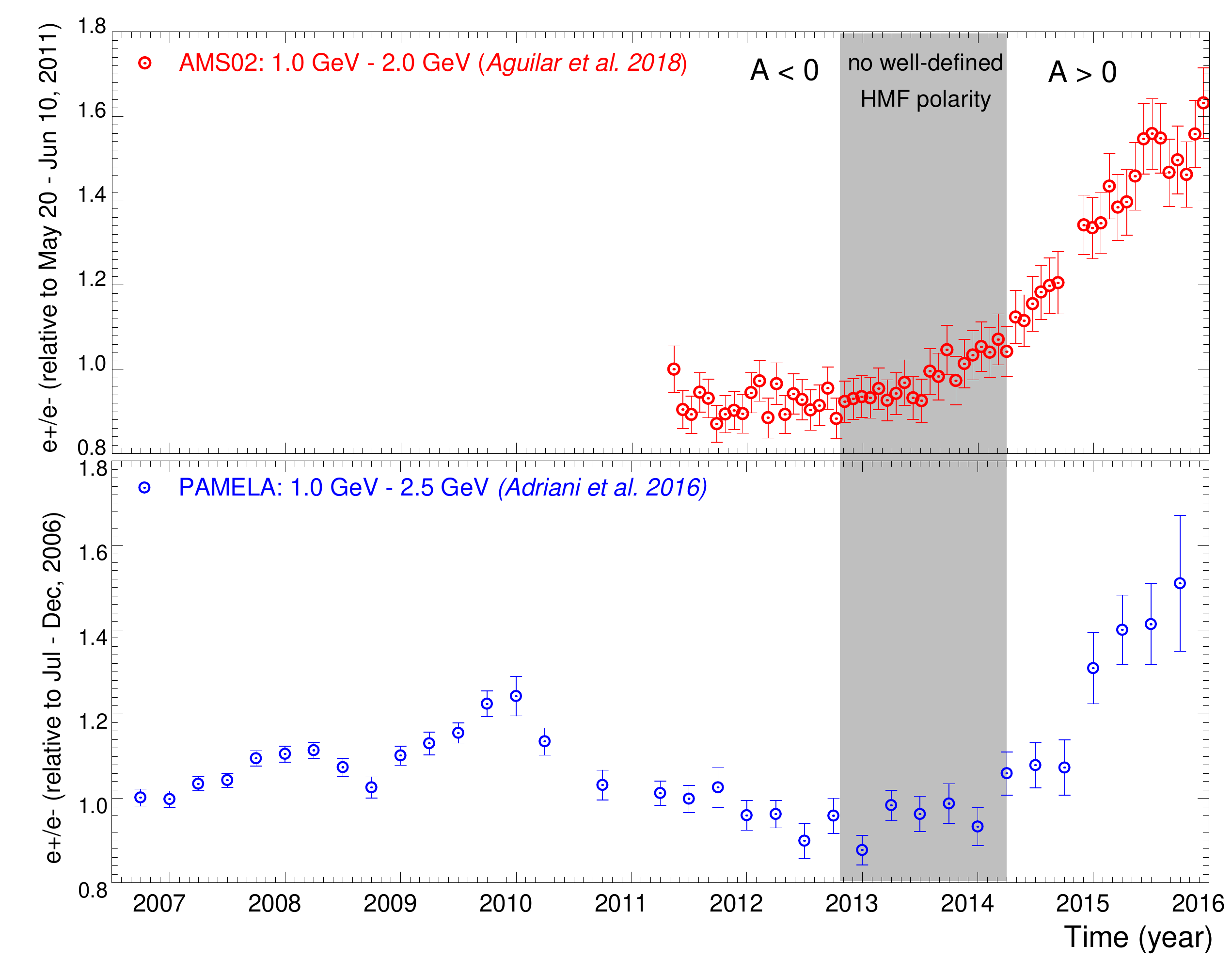}
\caption{Top panel: The ratio $e^{+}/e^{-}$ as observed by $\it{AMS}$02 from May 2011 to December 2015 for a KE range of 1.0 GeV to 2.0 GeV, relative to ratio for May 20 to June 10, 2011 (Bartel rotation: 2426; Aguilar et al. 2018b). Bottom panel: The ratio $e^{+}/e^{-}$ from July 2006 to December 2015, relative to July-December 2006, as measured by $\it{PAMELA}$ for a KE range of 1.0 GeV to 2.5 GeV (Adriani et al. 2016a). Shaded parts indicate the period of the HMF polarity reversal, with no well-defined HMF polarity; the polarity phase before this reversal is A$<$0, while A$>$0 afterwards. \label{fig1}}
\end{figure*}

In the bottom panel of Figure \ref{fig1} the three-month-average $e^{+}$/$e^{-}$ ratio as measured by $\it{PAMELA}$ from July 2006 to December 2015 is shown, relative to the period July - December 2006, for a range of 1.0 GeV - 2.5 GeV (adapted from Adriani et al. 2016a). The shaded portion indicates the HMF polarity reversal period treated as a phase with no well-defined HMF polarity. An increase of about 20\% in this ratio was observed up to the end of 2009, when it started to decrease continuously up to mid-2012, as solar activity had increased, then remaining almost the same for most of 2013, followed by an increase slowly up to the end-2014, with a sudden rise and progressive increase during 2015. The ratio ended up by being a factor of about 1.5 larger than in 2006. Adriani et al. (2016a) discussed this observed ratio, as well as two additional energy ranges in detail. The top panel of Figure \ref{fig1} shows the Bartel rotation averaged $e^{+}$/$e^{-}$ normalized to May 20 - June 10, 2011, measured by $\it{AMS}02$ from May 2011 to December 2015, over the energy range of 1.0 GeV - 2.0 GeV (adapted from Aguilar et al. 2018b). Similar to the $\it{PAMELA}$ observations, the ratio remains almost the same from May 2011 up to mid-2013, then started to increase, first modestly but then faster, to end up 60\% higher by the end of 2015. Not shown here is that this ratio then flatten off to remain almost the same from January 2016 to July 2017, marked as the end of the  report by Aguilar et al. (2018b), who also showed similar qualitative trends in this ratio over time up to about 10 GeV.    

As mentioned in the introduction, the prime objective of this numerical modeling is to reproduce the ratio $e^{+}$/$e^{-}$ as observed by $\it{PAMELA}$ from July 2006 to December 2015 and by $\it{AMS02}$ for the overlapping period from May 2011 to December 2015. In the process, we strive to understand the physics of how this charge-sign dependent effect had occurred and developed, especially over the HMF polarity reversal phase, using the mentioned 3D numerical modulation model. An important prerequisite for such detailed numerical modeling of full spectra is to specify the appropriate very local interstellar spectrum for each type of GCR particle, which is discussed next.

\section{Very Local Interstellar Spectra for Galactic Electrons and Positrons} \label{sec3}

As a mandatory initial condition, numerical modeling of the heliospheric modulation of GCRs requires to specify a galactic spectrum, more specifically a very local interstellar spectrum (VLIS), as an input spectrum, which is then modulated from the boundary of the simulated heliosphere, assumed to be the heliopause (HP) at 122 au, up to the Earth at 1 au. $\it{Voyager}$ 1 actually crossed the HP at a heliocentric distance of 122 au in August 2012 (Stone et al. 2013), with $\it{Voyager}$ 2 crossing the HP at 119 au in November 2018 (Stone et al. 2019). Beyond the HP these two probes are observing what can be called a very local interstellar spectrum for various GCR species, but their measurements cover a relatively narrow energy range, generally not higher than about 350 MeV/nuc, but only between 2.7 MeV and 74 MeV for electrons (Cummings et al. 2016). These observations nevertheless have had a significantly influence on improving the confidence levels of what are used nowadays for the HP position and for VLIS, especially important to electrons (and positrons) because the shape (and slope) of their VLIS at these low kinetic energies may be preserved even up to Earth (Nndanganeni \& Potgieter 2018) whereas for protons and GCR nuclei the adiabatic energy losses completely alter the shape of their VLIS by the time they are observed at the Earth; see Vos \& Potgieter (2016); Ngobeni et al. (2020). 

In previous modulation modeling studies the $\it{Voyager}$ observations together with the $\it{PAMELA}$ measurements of GCRs at much higher kinetic energies were used effectively to construct new VLIS for galactic protons (Potgieter et al. 2014; Vos \& Potgieter 2015) and galactic electrons (Potgieter et al. 2015; Potgieter \& Vos 2017) by finding the best statistical fit to the two mentioned data sets. For positrons, however, a different approach must be taken, utilizing the availability of sophisticated and comprehensive galactic propagation models (GALPROP) along with the 3D numerical modulation model to construct new self-consistent VLIS for GCRs as acknowledged and described in detail by Bisschoff et al. (2019); see also Bisschoff \& Potgieter (2014, 2016) and references therein. These self-consistent VLIS based on GALPROP models for different GCR particles together with the mentioned observations, where available, have been used extensively in recent modeling studies of GCR protons (Aslam et al. 2019a; Ngobeni et al. 2020), electrons and positrons (Aslam et al. 2019b), and helium (Marcelli et al. 2020; Ngobeni et al. 2020). Positron and electron VLIS computed with this approach are discussed next.

\begin{figure}[!htp]
\plotone{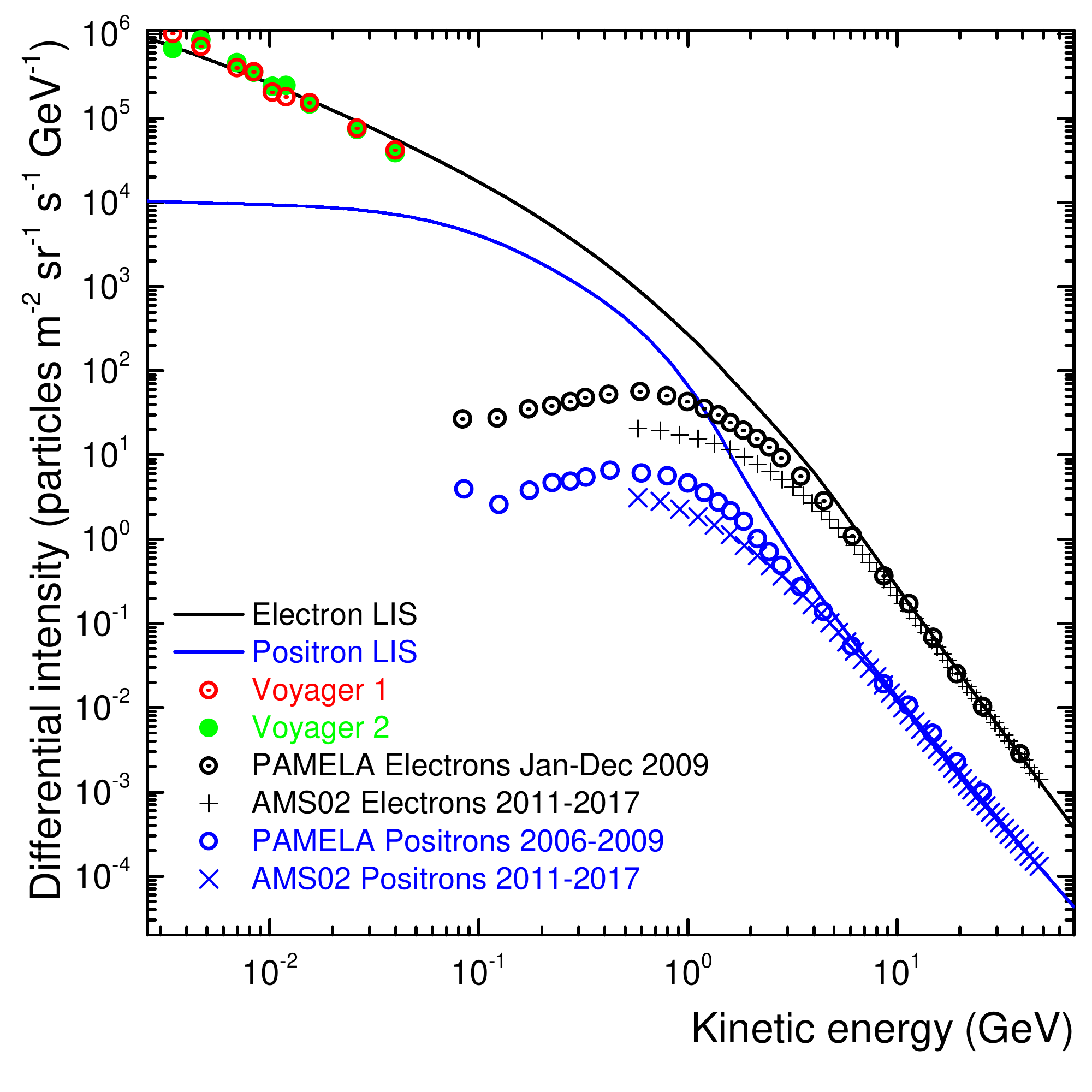}
\caption{Black solid line depicts the VLIS for GCR electrons as computed with GALPROP and 'tuned' to $\it{Voyager}$ 1 (Stone et al. 2013; Cummings et al. 2016) and $\it{Voyager}$ 2 observations (Stone et al. 2019) at low kinetic energies and $\it{PAMELA}$ (Adriani et al. 2015) and $\it{AMS}02$ observations (Aguilar et al. 2018b) at high kinetic energies, averaged for the periods as indicated. The blue solid line shows the positron VLIS constructed and used by Aslam et al. (2019a, b) by modifying the GALPROP computed VLIS using $\it{PAMELA}$ (Munini 2015) and $\it{AMS}02$ observations (Aguilar et al. 2018b), averaged for the periods as indicated. The procedure validates the VLIS at very low and high energies, respectively. \label{fig2}}
\end{figure}

Figure \ref{fig2} shows the VLIS for $e^{-}$ and $e^{+}$ computed with the web version of the GALPROP code (see Moskalenko \& Strong 1998; Vladimirov et al. 2011, and references therein) as described by Bisschoff et al. (2019). Essentially, the appropriate physics in GALPROP is adjusted until the observed $\it{Voyager}$ electron intensity levels beyond the HP are obtained and for high KE observations from $\it{PAMELA}$ where solar modulation is considered negligible. Only the latter can be done for $e^{+}$ so that the shape of their VLIS at low energies is based on what GALPROP produced. The observations displayed in this figure are: Electrons from $\it{PAMELA}$ from January - December 2009 (Adriani et al. 2015) and from $\it{AMS}02$ from May 2011 - May 2017 (Aguilar et al. 2018b), and from $\it{Voyager}$ 1 (Gurnett et al. 2013; Stone et al. 2013; Webber \& McDonald 2013; Cummings et al. 2016) and $\it{Voyager}$ 2 (Stone et al. 2019; Krimigis et al., 2019); positron observations are from $\it{PAMELA}$ (Munini 2015) and $\it{AMS}02$ (Aguilar et al. 2018b). It should be noted that the $\it{Voyager}$2 observations also contains positrons. The black solid line in this figure represents the computed VLIS for electrons using observations from $\it{PAMELA}$ and from $\it{AMS}02$ and observations from $\it{Voyager}$ 1 and $\it{Voyager}$ 2. For our purpose, these data sets serve as validation of the applied VLIS. 

Concerning the positron VLIS, Aslam et al. (2019a) found that when using the GALPROP computed $e^{+}$ LIS, they could not reproduce the $e^{+}$s spectra from $\it{PAMELA}$ for the 2006 -2009 period when using for $e^{+}$ the exact same modulation parameter values as for $e^{-}$ modulation for this period. So this computed $e^{+}$ VLIS was empirically adjusted to improve the reproduction of the mentioned data set. This is shown by the blue solid line in Figure \ref{fig2}, and is also used in this study. This particularly modified $e^{+}$ VLIS was used by Aslam et al. (2019a) to reproduce the 6-month averaged $\it{PAMELA}$ $e^{+}$s for July 2006-December 2009 and the Bartel rotation averaged $\it{AMS}02$ $e^{+}$s for May 2011-May 2017 (Aslam et al. 2019b). 

\section{Numerical Simulations of Galactic Electron and Positron Spectra} \label{sec4}

A sophisticated and comprehensive three-dimensional (3D) steady-state model, which is based on the numerical solution of Parker's transport equation (Parker 1965), is used to compute the differential intensity of $e^{-}$ and $e^{+}$ over a KE range 1.0 MeV to 70 GeV, from the Earth (1 au) and at radial distances up to the HP at 122 au. Here, the VLIS is specified as an initial condition in the numerical model. Any modulation beyond the HP, as explained by Luo et al. (2015, 2016), is neglected in this approach. 

Parker's transport equation (TPE) is described as:  
\begin{equation}
\frac{\partial f}{\partial t} = - \vec{V}_{sw} \cdot \nabla \it{f} - \langle \vec {v}_{D} \rangle \cdot \nabla \it{f} + \nabla \cdot (\bf{K}_{s} \cdot \nabla \it{f}) + \frac {1}{3} (\nabla \cdot \vec{V}_{sw}) \frac {\partial f} {\partial  ln  p} \label{Eq1}
\end{equation}

where $f (\vec {r}, p, t)$ is the particle distribution function, $\it{p}$ is momentum, $\it{t}$ is time, and $\vec {r}$ is the vector position in 3D, with the three coordinates $\it{r}$, $\theta$, and $\phi$ specified in a heliocentric spherical coordinate system where the equatorial plane is at a polar angle of $\theta = 90^{\circ}$.
 
The terms shown on the right-hand side of this equation represent: (1) The outward convection caused by the expanding solar wind with velocity ($\vec {V}_{sw}$), (2) the averaged particle drift velocity $\langle \vec {v}_{D} \rangle$ (pitch angle averaged guiding center drift velocity), which is described by
\begin{equation}
\langle \vec {v}_{D} \rangle = \nabla \times K_{D} \frac{\vec B}{B}
\label{Eq2}
\end{equation}

where $K_{D}$ is the generalized drift coefficient, and $\vec{B}$ is the HMF vector with magnitude $\it{B}$. (3) Spatial diffusion caused by scattering of GCRs, where $\bf{K}_{s}$ is the symmetry diffusion tensor, and (4) adiabatic energy change, which depends on the sign of the divergence of $\vec {V}_{sw}$. If $\nabla \cdot \vec{V}_{sw}>0 $, adiabatic energy loss occurs as is the case in most of the heliosphere, except inside the heliosheath where we assume that $\nabla \cdot \vec{V}_{sw} = 0$; see also Langner et al. (2006). 

As mentioned above, there are the four major physical processes which GCR particles undergo when they enter and travel through the heliosphere up to the Earth. All these physical processes play an important role in modulating GCR particles, but their relative importance and dominance vary according to the solar activity cycle and depend also on the solar magnetic polarity cycle, and of course on the GCR species. Potgieter et al. (2014, 2015) and Aslam et al. (2019a) described the 3D numerical modulation model used for this study in detail, and the reviews from Potgieter (2013, 2014b, 2017) also explain the underlying theory for the global modulation of GCRs,  especially during a relatively quiet heliosphere as has been observed for the past two solar cycles.

We note that this steady-state 3D model is not suitable to study GCR events shorter than one solar rotation period. The reason for keeping steady-state is that the Alternating Direct Implicit numerical method is not stable when a fifth numerical dimension is incorporated into the numerical scheme (that is, three spatial, one for energy dependence, and one for time dependence). If all three spatial dimensions are used, the time dependence part must be sacrificed. In 2D models, such as used by le Roux \& Potgieter (1995), Langner et al. (2006), Ngobeni et al. (2011) and Manuel et al. (2014), the azimuthal dependence is omitted. A full 3D study of propagating transient events such as Forbush decreases (FDs) or GCR variations caused by Corotating Interaction regions (CIRs) need to be done using the Stochastic Differential Equation (SDE) approach (Zhang, 1999, Bobik et al. 2016, Strauss \& Effenberger 2017) of solving the TPE, as applied to different aspects in modulation modeling as reported by e.g. Luo et al. (2017, 2018, 2020).
 
The functional forms of the three diffusion coefficients and the drift coefficient as used in the numerical model, and how they change with time, to illustrate the underlying physics, are given and discussed in what follows. 

\subsection{Calculation and Selection of Intrinsic Modulation Parameters} \label{sec4.1}

The use of this 3D steady-state model makes the determination of representative values for time-varying modulation parameters a challenging task. Solar activity parameters used in the modeling such as the tilt angle $\alpha$ of the HCS and the magnitude $B$ of the HMF at the Earth, change continuously over time. In order to set up representative modulation conditions as input for the numerical model, appropriate moving averages of $\alpha$ and $B$ were used and a similar approach was applied to the GCR data sets, both for $\it{AMS}02$ (Corti. et al. 2019) and $\it{PAMELA}$ observations (Vos \& Potgieter 2015; Aslam et al. 2019a; Ngobeni et al. 2020), which are not always averaged over the same period, depending on the time resolutions of these observations.

In the model, the fast solar wind in the heliospheric polar regions has a maximum speed of $\approx750$ km s$^{-1}$, while the slow solar wind in the equatorial region has an average speed of $\approx430$ km s$^{-1}$ for the solar minimum period of 2006 - 2009. This assumption is based on $\it{Ulysses}$ observations for the latitudinal dependence of the solar wind speed reported by McComas et al. (2008). This dependence decreases gradually as solar activity increases so that no clear latitudinal dependence occurs for solar maximum conditions.

During minimum to moderate solar activity the HCS is mostly confined to the ecliptic region (within $\approx$ 30$^{\circ}$), thus remaining in the slow solar wind region, so the slow solar wind speed is used to calculate the time the changes in $\alpha$ takes to travel from the Sun to the HP at 122 au. However, the HMF is not confined to the ecliptic region, so a weighted average of both slow and fast solar wind speed is used to calculate the time taken to reach the HP from the Sun for the solar minimum period of 2006-2009. As the latitudinal dependence of the solar wind  decreases with increasing solar activity, only the slow solar wind speed is used to calculate the propagation time of changes in the HMF from the Sun to the HP, that is, for moderate to maximum solar activity. The propagation time calculated for $\alpha$ is $\approx$15-months (17-solar rotations) and for the HMF it is $\approx$10-months (11-solar rotations) for the solar minimum of 2006-2009, and $\approx$15-months (17-solar rotations) for the moderate to maximum solar activity phases, 2010-2016. The moving averages of both $\alpha$ and $B$ over the propagation time are used as intrinsic parameters in this modeling approach, similar to what were used in numerical studies by Potgieter et al. (2014, 2015), Vos \& Potgieter (2015, 2016) and Ngobeni et al. (2020). 

\begin{figure*}[!htp]
\plotone{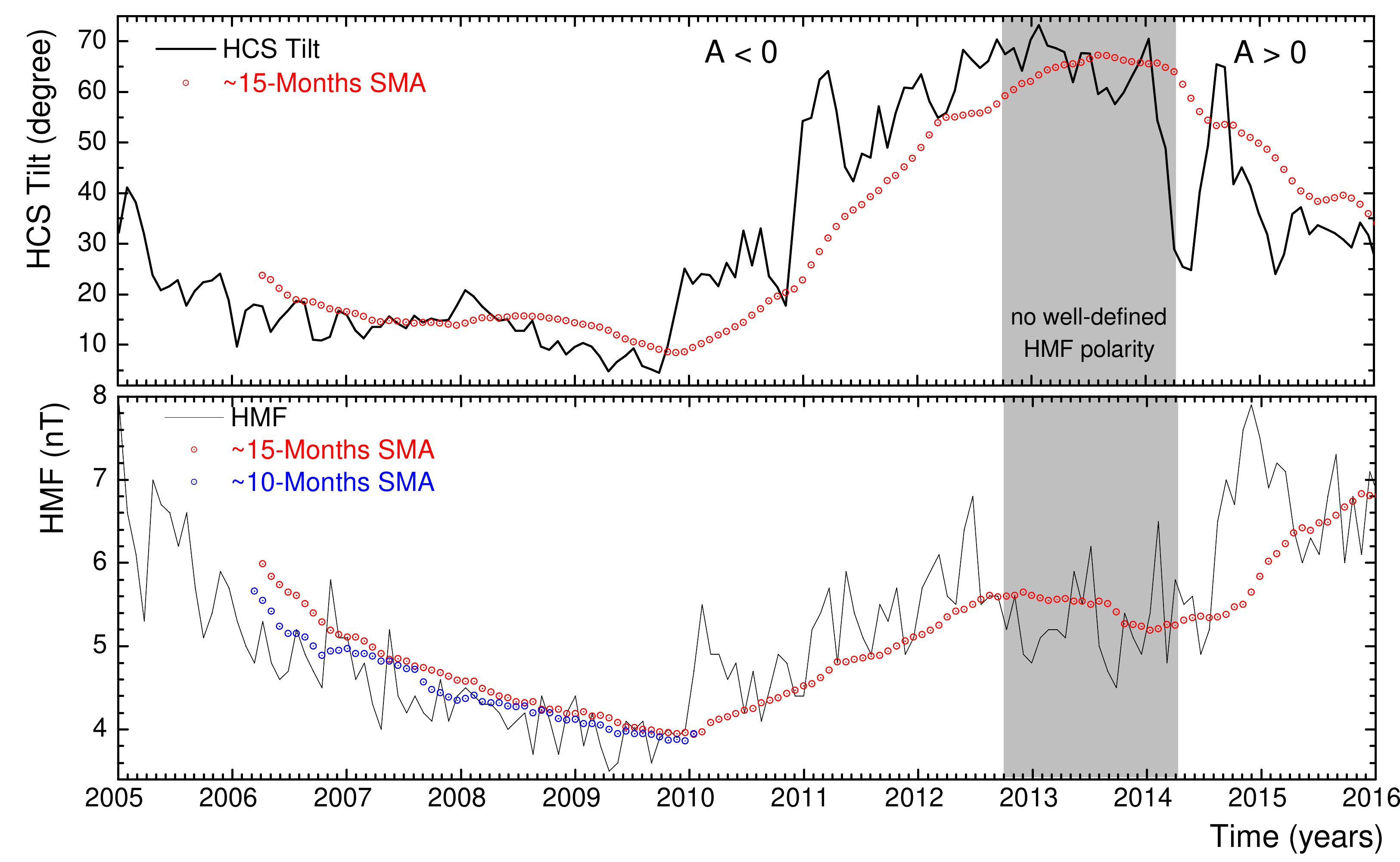}
\caption{Top panel: Tilt angle $\alpha$ of the HCS (black line) at the Earth from January 2005 to December 2015 taken from http://wso.stanford.edu, along with 17-Carrington rotation ($\approx$ 15-months) moving averages (red dots). 
Bottom Panel: Magnitude $B$ of the HMF at the Earth (black line) for the same period taken from http://omniweb.gsfc.nasa.gov, along with 11-Bartel rotation ($\approx$ 10-months) moving averages (blue dots; used only for 2006-2009), and  17-Bartel rotation ($\approx$ 15-months) moving averages (red dots). Shaded portions indicate the period of the polarity reversal of Sun's magnetic field which is considered as a period of no well-defined HMF polarity. \label{fig3}}
\end{figure*}

Figure \ref{fig3} shows both the Carrington rotation average $\alpha$ as adopted from http://wso.stanford.edu and Bartel rotation average $B$ as adopted from http://omniweb.gsfc.nasa.gov for January 2005 to December 2015, and the continuous moving averages over the mentioned propagation times as applied in this study.
      
$\it{Voyager}$ 2 measurements showed a sudden decrease in the solar wind speed around $\approx$ 84 au which corresponds to the crossing of the termination shock (TS). Studies by Richardson \& Wang (2011) using these observations reported that the position of the TS is moving inward and outward in response to changes in the dynamic pressure of the solar wind. This dynamic nature of the TS caused by solar activity variations is also incorporated in the model because the changing width of the inner heliosheath affects the modulated intensities at the Earth, see Langner \& Potgieter (2005), Manuel et al. (2014), Vos \& Potgieter (2016) and Aslam et al. (2019a). In the model, the position of the TS is changed from 88 au to 80 au for the solar minimum activity period of 2006-2009 and then back from 80 au to 88 au as solar activity increased gradually to maximum from 2010 to 2016. 

A few remarks about Figure \ref{fig3} and what it means in terms of using them as time varying proxies for solar activity: The HCS $\alpha$ was below 30$^{\circ}$ from 2006 to 2009, with a modest increase during 2008, to become below 10$^{\circ}$ for almost one year until the end-2009. It then increased rapidly in 2010 and in 2011 even beyond 60$^{\circ}$, indicating that solar maximum modulation conditions had been reached within two years after 2009. These high levels had been maintained without much fluctuation until mid-2014 indicating a clear and stable solar maximum phase, later in 2014 $\alpha$ displayed large fluctuations, coming down to 30$^{\circ}$ late in 2015, indicating that a period of relatively calm modulation conditions have been reached already since the beginning of 2016 until the present time. (For details, see the HCS tilt observations from Wilcox solar observatory http://wso.stanford.edu). The HMF magnitude $B$ at Earth reached extraordinary low levels during 2008 and 2009. Similar to $\alpha$, it had increased rapidly in 2010, followed by a gradual increase with many fluctuations towards maximum around mid-2012. But then $B$ had a sudden drop to below even 5 nT just after reaching the maximum and then had stayed below about 6 nT for the whole polarity reversal period. Later in 2014, it displayed a sudden increase to 8 nT with many fluctuations. This behavior was probably caused by the increase in the number of Coronal Mass Ejections (CMEs: with a significant release of plasma and accompanying magnetic field from the solar corona, see the list of ICMEs, http://www.srl.caltech.edu/ACE/ASC/DATA/level3/icmetable2.htm). Evidently, the use of moving averages smooth out these large fluctuations, which may not always be optimal.

\subsection{Comparison of modeling results with observations} \label{sec4.2}     

In order to achieve the primary objective of this work, $e^{-}$ and $e^{+}$ energy spectra measured by $\it{PAMELA}$, averaged over 6- months from July 2006 to December 2009 (Adriani et al. 2015; Munini 2015), and $\it{AMS}02$, averaged over a Bartel rotation period from May 2011 to December 2015 (Aguilar et al. 2018b) were utilized. Instead of considering all Bartel rotation averaged spectra measured by $\it{AMS}02$, we choose only two sets for each year, the first one corresponds to the middle of the year and the second one corresponds to the end of the year and then generalized for each 6-months. The spectra selected from $\it{AMS}02$ specifically are May 20 - June 10, 2011 (Bartel rotation 2426; observation starts from May 20, 2011), November 20 - December 16, 2011 (Bartel rotation 2433), May 27 - June 22, 2012 (Bartel rotation 2440), December 02 - December 28, 2012 (Bartel rotation 2447), May 13 - June 08, 2013 (Bartel rotation 2453), November 18 - December 146, 2013 (Bartel rotation 2460), May 26 -June 21, 2014 (Bartel rotation 2467), December 01 - December 27, 2014 (Bartel rotation 2474), May 12 - June 07, 2015 (Bartel rotation 2480), and November 17 - December 13, 2015 (Bartel rotation 2487).   
   
First, we compute full $e^{-}$ spectra, with the purpose of reproducing the mentioned $\it{PAMELA}$ spectra, using for every new simulation run the calculated averaged values for $\alpha$ and $B$ as input for the model, and carefully adjusting the diffusion and drift coefficients accordingly. Then, using the exact same modulation parameters we reproduced $e^{+}$ spectra of the same period, similar to what was done by Aslam et al. (2019a). As such, the only two differences between $e^{-}$ and $e^{+}$ modulation are their respective VLIS and the particle drift that they experience. Repeating the procedure, we next reproduced the selected $e^{-}$  and $e^{+}$ spectra from $\it{AMS}02$ using the exact same set of modulation parameters for both GCR species.

\begin{figure*}[!htp]
\plotone{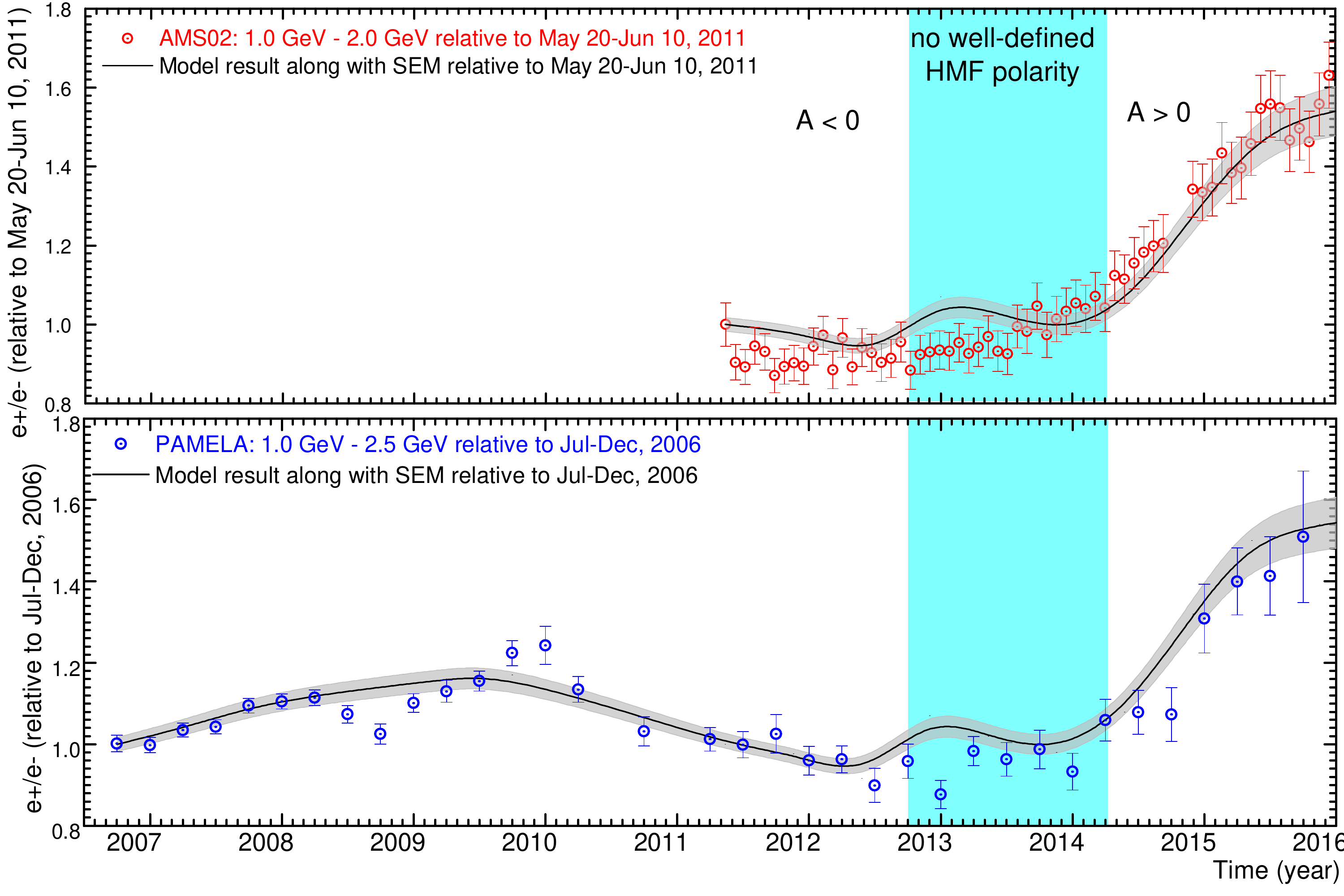}
\caption{Top panel: Modeling of $e^{+}$/$e^{-}$ (solid line, with standard error of mean $SEM$ = standard deviation/$\sqrt n$) is shown in direct comparison with $\it{AMS}02$ observations for the KE range of 1.0 GeV - 2.0 GeV (red dots; Aguilar et al. 2018b). These observations are averaged over Bartel rotations for May 2011 - December 2015. Both ratios are normalized with respect to May 2011.
Bottom Panel: Observed $e^{+}$/$e^{-}$ by $\it{PAMELA}$ over a KE range of 1.0 GeV - 2.5 GeV (blue dots; Adriani et al. 2016a), averaged over 3-months for July 2006 - December 2015, is compared with the computed $e^{+}$/$e^{-}$ over the KE range of 1.0 GeV - 2.0 GeV (solid line, standard error of mean). Here both ratios are normalized with respect to July- December 2006. As before the shaded regions indicate the period without a well-defined HMF polarity.\label{fig4}}
\end{figure*}

In the bottom panel of Figure \ref {fig4} the $\it{PAMELA}$ observed $e^{+}$/$e^{-}$ reported by Adriani et al. (2016a) for July 2006 - December 2015 for the KE range of 1.0 GeV - 2.5 GeV is compared with the computed $e^{+}$/$e^{-}$ for 1.0 GeV - 2.0 GeV; both ratios are normalized with respect to July - December 2006. The shaded region indicates the polarity reversal phase of the HMF, without a well-defined HMF polarity; the period before this reversal is A$<$0, and the period afterwards is A$>$0. Evidently, the modeling ratio follows the same trend than the observed ratio, that is, increasing gradually from 2006 to a maximum around the end of 2009 for the A$<$0 cycle, then decreasing gradually to reach at minimum level in 2012, just before and during the first phase of the reversal period; the ratio increasing somewhat during the reversal period but remains essentially unchanged; after the reversal period the ratio increases significantly and progressively until it levels off during 2015, as observed. In the top panel of Figure \ref {fig4}, the $\it{AMS}02$ observed $e^{+}$/$e^{-}$ for May 2011 - December 2015 over the KE range of 1.0 GeV - 2.0 GeV (Aguilar et al. 2018b) is compared with the computed ratio; both ratios are normalized with respect to May 2011. Again the model follows the trend of these observations, giving the ratio clearly higher in the A$>$0 polarity cycle than during the A$<$0 cycle. Comments and further elaborations on these results are given in Section \ref{sec5}.

The following sections elaborate on the modulation parameters used to reproduce the observed $\it{PAMELA}$ and $\it{AMS}02$ $e^{+}$/$e^{-}$, focusing on how the physics related to the drift and diffusion coefficients over time changes from the solar minimum of 2006-2009 to solar maximum in 2015, including the HMF polarity reversal phase of November 2012 to March 2014. 
 
\subsection{Drift Coefficient over Time} \label{sec4.3} 
    
The expression for the drift coefficient $K_{D}$ in Equation (\ref {Eq2}) can be rewritten as
\begin{equation} 
K_{D} = \frac {\beta P} {3B_{m}} f_{D} = \frac {\beta P} {3B_{m}} \Bigg[ \frac {(\omega \tau)^{2}}{1+(\omega \tau)^{2}}\Bigg] \label{Eq3}
\end{equation}

where, $B_{m}$ is the magnitude of the Smith-Bieber modified (Smith \& Bieber 1991) Parker-type HMF ($\vec B_{m}$), $\beta$ = $v/ c$ is the ratio of particle speed to the speed of light, $P$ is particle rigidity, and $\omega$ is the particle gyro-frequency with $\tau$ the average time between the scattering of GCR particles in the turbulent HMF.

The term $f_{D}$ = $\Bigg[ \frac {(\omega \tau)^{2}}{1+(\omega \tau)^{2}}\Bigg]$ in the Equation (\ref {Eq3}) is called the drift reduction factor and is determined by how diffusive scattering is described; if $f_{D}$ = 0, then $K_{D}$ and therefore the drift velocity $\langle v_{D} \rangle$ becomes zero so that drift effects will vanish from the modulation model to produce non-drift solutions; if $f_{D}$ = 1, drift is at a maximum, so that $K_{D}$ then has the weak scattering value because $\omega \tau$ becomes much larger than 1.0, in the order of 10 (see e.g., Ngobeni et al. 2015, and references there-in). Theoretically, this gives the largest possible value for $K_{D}$, for a given particle rigidity and $B$ value, as found for the unmodified Parker HMF (see illustrations by Raath et al. 2016). If this is the case, drift effects in GCR modulation are predicted to be very large and dominant as found in original numerical models by Jokipii \& Kopriva (1979; where the $A>0$ and $A<0$ computed spectra were the opposite of what was observed), K\'{o}ta \& Jokipii (1983), Potgieter \& Moraal (1985), to mention only a few.

This function $f_{D}$ can be used to adjust the rigidity dependence of $K_{D}$, which is the most effective direct way of suppressing drift effects at low rigidities as required by $\it{Ulysses}$ observations of latitudinal gradients (see reviews by Heber \& Marsden 2001; Heber \& Potgieter 2006; Heber 2013) so that Equation (\ref {Eq3}) becomes:

\begin{equation}
K_{D} = \frac {\beta P} {3B_{m}} f_{D} = K_{A0}  \frac {\beta P} {3B_{m}} \frac {(P/P_{A0})^{2}}{1+(P/P_{A0})^{2}} \label{Eq4}
\end{equation}           

Here, $K_{A0}$ is dimensionless, and could be ranging from 0.0 to 1.0. In this study, we keep $K_{A0}$ = 0.90 and $P_{A0}$ = 0.90 GV for the period from July 2006 to June 2010, which means particle drift is at a 90\% level of a theoretical maximum during this unusual solar minimum period, but below $\approx$ 1.0 GV (determined by the assumed value for $P_{A0}$), the drift coefficient is reduced with respect to the weak scattering approach. For details on such a modeling approach, see also Ngobeni et al. (2015); Nndanganeni \& Potgieter (2016, 2018), and Aslam et al. (2019a); for latest theoretical advances on drift reduction, see e.g. Engelbrecht et al. (2017). 

However, for the reproduction of the $e^{+}$/$e^{-}$ during the modulation period after June 2010, $K_{A0}$ had to decrease gradually and systematically for this A $<$ 0 cycle from 0.90 to 0.0, which is required for the polarity reversal period without a well-defined polarity. After this reversal period, with A $>$ 0$, K_{A0}$ is changing gradually from 0.0 to 0.65 (for December 2015) as solar activity decreases. When $K_{A0}$ = 0, also $K_{D}$ = 0, which means that for the reversal period from November 2012 - March 2014 no drifts are present in the model on a global scale. The latter implies that perhaps some localized drifts may be present somewhere in the real heliosphere, but with no global modulation consequences so that the effect of global drifts can be neglected for this polarity reversal phase during solar maximum activity. This means that any charge-sign dependence in solar modulation also dissipates.
    
\begin{figure}[!htp]
\plotone{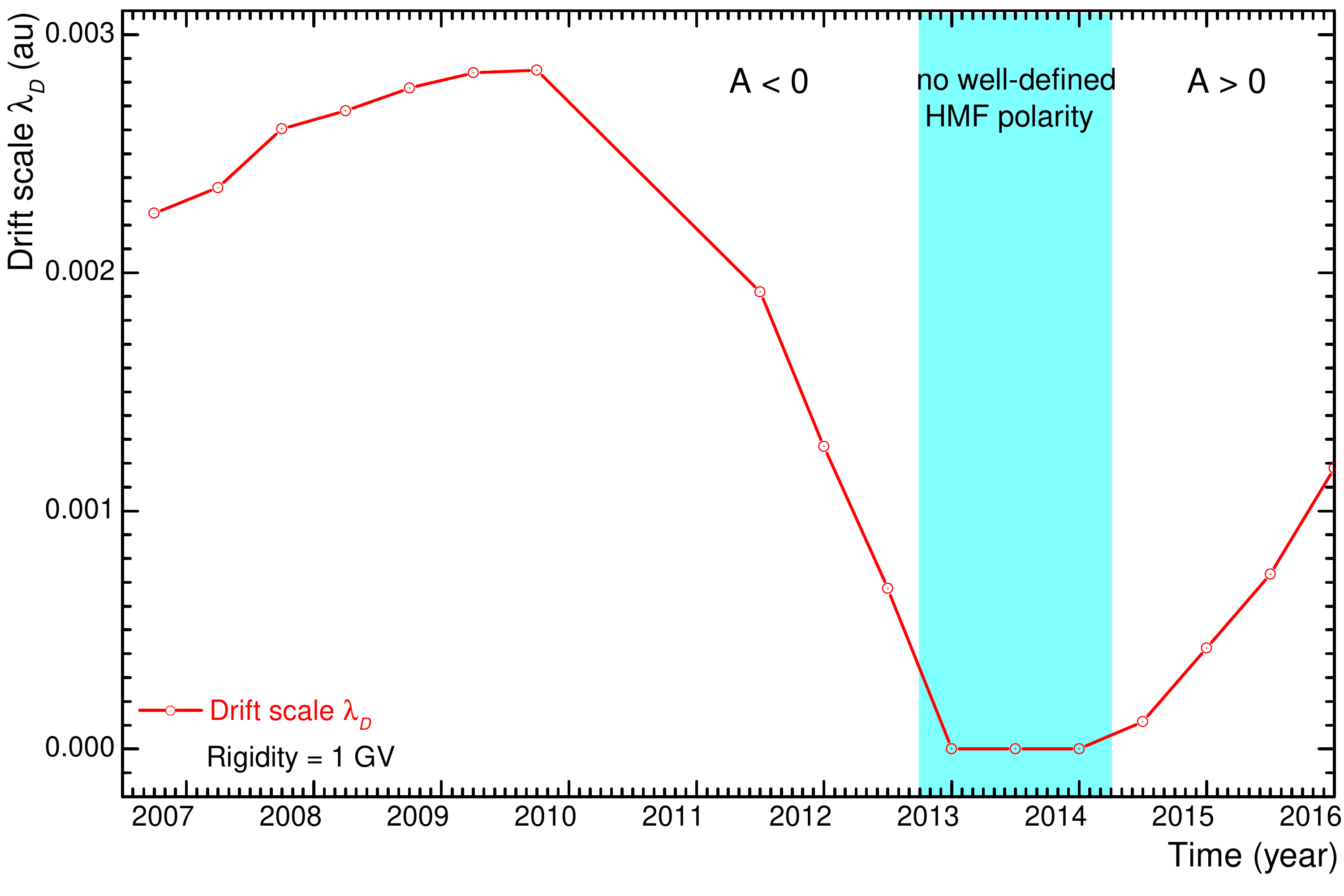}
\caption{Changes in the drift scale $\lambda_D$ (in au) based on $K_{D}$ as in Equation (\ref{Eq4}) for 1.0 GV electrons and positrons from 2006 July - December 2015 as required to produce the modeling results shown in Figure \ref{fig4}. 
Points for January 2010 to April 2011 were obtained through extrapolation, as explained in the text, because no published electron and positron data were available then. 
\label{fig5}}
\end{figure}

The drift scale $\lambda_D$ = $(3/\beta c$)$K_D$ is calculated for 1.0 GV electrons and positrons using $K_{D}$ as in Equation (\ref {Eq4}) for each selected period from July 2006 - December 2015 and is shown in Figure \ref{fig5}. The advantage of plotting $\lambda_D$ is that it can be expressed in units of au (instead of $m^{2}/s$ as for $K_{D}$). For 2006 to 2009, although $K_{A0}$ = 0.90 and $P_{A0}$ = 0.90 GV were kept unchanged in Equation (\ref {Eq4}), $\lambda_D$ ($K_{D}$) increased gradually to reach a maximum value around the end of 2009 because of the decrease in $B$. The value of $P_{A0}$ = 0.90 GV was kept unchanged for the period from 2010 to December 2015 but in addition to the changes in $K_{D}$ caused by $B$ over this period, also the value of $K_{A0}$ was changed progressively as solar activity had progressed. Obtaining $\lambda_D$ ($K_{D}$) in this manner is directly based on reproducing with the model electron and positron observations from $\it{PAMELA}$ for July 2006 -December 2009, and for $\it{AMS}02$ for May 2011 -December 2015. For the period in between (January 2010 - May 2011; for when no published data are available), $\lambda_D$ is extrapolated using only averaged $B$ values, while keeping the same value of the previous period for $K_{A0}$. From this figure follows that $\lambda_D$ ($K_{D}$) reaches a maximum value around the end of 2009, then gradually decreases as solar activity increases to become eventually negligible during the polarity reversal phase. After the polarity reversal, it recovers systematically, gaining until the end of 2015 roughly up to the 50\% of the 2009 level. See also Aslam et al. (2019a).

\subsection{Diffusion Coefficients over Time} \label{sec4.4}   

The expression for the diffusion coefficient parallel to the average background HMF, as used in this modeling effort, is given by:

\begin{equation}
K_{\parallel} = (K_{\parallel})_{0} \beta \Bigg(\frac {B_{0}}{B}\Bigg) \Bigg(\frac {P}{P_{0}}\Bigg)^{c_{1}} \left[ \frac {\Bigg(\frac {P}{P_{0}}\Bigg)^{c_{3}} + \Bigg(\frac {P_{k}}{P_{0}}\Bigg)^{c_{3}}}{ 1+ \Bigg(\frac {P_{k}}{P_{0}}\Bigg)^{c_{3}}} \right]^{\frac {c_{2 \parallel} - c_{1}}{c_{3}}}
\label{Eq5}
\end{equation}

where $(K_{\parallel})_{0}$ is a scaling constant in units of $6\times10^{20}$ cm$^{2}$s$^{-1}$, with the rest of the equation written to be dimensionless with $P_{0}$ = 1.0 GV, and $B_{0}$ = 1.0 nT (in order to preserve the units in cm$^{2}$ s$^{-1}$). Here $c_{1}$ is a power index that may change with time if required; $c_{2 \parallel}$ and $c_{2 \perp}$ (see Equation \ref{Eq6}) together with $c_{1}$ determine the slope of the rigidity dependence, respectively, above and below a rigidity with the value $P_{k}$ which may change with time if required, $c_{3}$ determines the smoothness of the transition. The rigidity dependence of $K_{\parallel}$ is thus a combination of two power laws; $P_{k}$ determines the rigidity where the break in the power law occurs and the value of $c_{1}$ determines the slope of the power law at rigidities below $P_{k}$. How the rigidity dependence changes with time will be shown later in particular. In addition, the value of $(K_{\parallel})_{0}$ was changed from 32.13 in 2006 to 34.29 in 2009 and 31.38 in 2015.

The radial dependence of the diffusion coefficients in the inner heliosphere (less than 5 au) was adjusted according to Vos \& Potgieter (2015, 2016), and followed by Aslam et al. (2019a), in order to reproduce the radial gradients as observed by $\it{Ulysses}$ (Gieseler \& Heber 2016) which requires an increase in the diffusion coefficients at these radial distances.

\begin{figure}[!htp]
\plotone{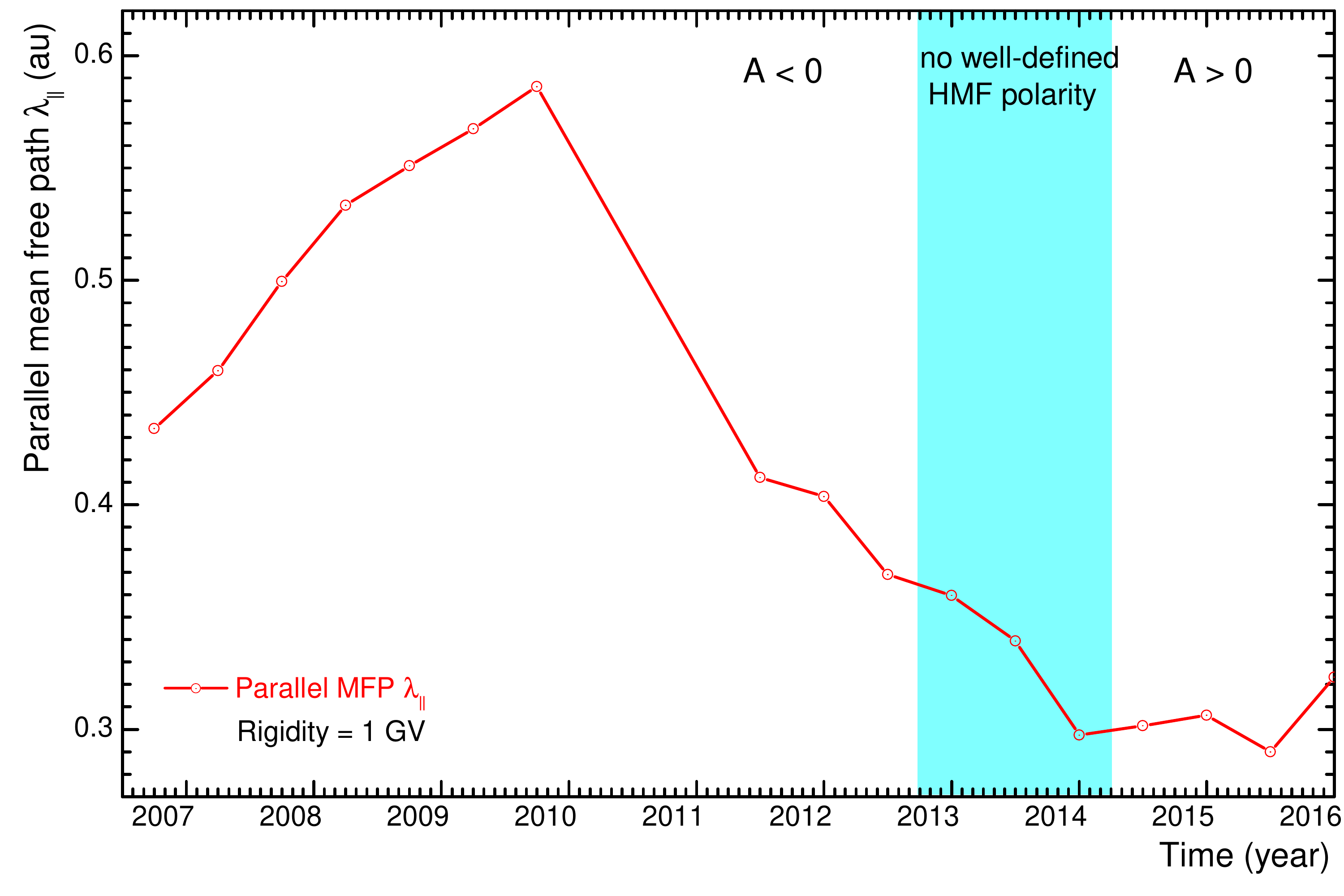}
\caption{Computed parallel mean free path ($\lambda_{\parallel}$) as used in the model for 1.0 GV electrons and positrons from 2006 July to December 2015. As for the drift scale, the MFP for January 2010 to April 2011 were obtained through extrapolation.
 \label{fig6}}
\end{figure}

The relation between diffusion coefficients ($K$) and their corresponding mean free paths (MFPs; $\lambda$) in general is: 
$K$ = $\lambda$($v/3$), where $v$ is the speed of the particles. The diffusion coefficient parallel to the average background HMF, $K_{\parallel}$, was calculated at 1.0 GV (as an example) for electrons and positrons, using Equation (\ref {Eq5}), from mid-2006 to the end of 2015. This calculated time variation from July 2006 to December 2015 is shown in Figure \ref{fig6} but plotted as the corresponding parallel MFP ($\lambda_{\parallel}$). The advantage of plotting MFPs instead of diffusion coefficients is that they can be given in units of au and are independent of the particle's speed ($\beta$ value).

Perpendicular diffusion in the radial direction is assumed to scale spatially similar to Equation (\ref {Eq5}) but with a different rigidity dependence at higher rigidities,

\begin{equation}
K_{\perp r} = 0.02 (K_{\parallel})_{0}  \beta  \Bigg(\frac {B_{0}}{B}\Bigg) \Bigg(\frac {P}{P_{0}}\Bigg)^{c_{1}} \left[ \frac {\Bigg(\frac {P}{P_{0}}\Bigg)^{c_{3}} + \Bigg(\frac {P_{k}}{P_{0}}\Bigg)^{c_{3}}} { 1+ \Bigg(\frac {P_{k}}{P_{0}}\Bigg)^{c_{3}}} \right]^{\frac {c_{2 \perp}- c_{1}}{c_{3}}}
\label{Eq6}
\end{equation}

Using $K_{\perp r}$ = 0.02 $K_{\parallel}$ in general is a widely used and reasonable assumption (e.g. Giacalone et al. 1999); in case of electrons and positrons, the value of c$_{1}$ = 0.0. (See Nndanganeni \& Potgieter 2016 for illustrations of the modulation effect on electrons when changing this ratio). 

The polar perpendicular diffusion ($K_{\perp \theta}$), on the other hand is more complicated, with consensus that $K_{\perp \theta}$ $>$ $K_{\perp r}$ away from the equatorial regions as discussed and motivated by Potgieter (2000) and Potgieter et al. (2014). 
It is assumed to be given by

\begin{equation}
K_{\perp \theta} = 0.02 K_{\parallel} f_{\perp \theta} = K_{\perp r} f_{\perp \theta}
\label{Eq7}
\end{equation}
with
\begin{equation}
f_{\perp \theta} = A^{+} \mp A^{-} tanh[8(\theta_{A} - 90^{\circ}) \pm \theta_{F}].
\label{Eq8}
\end{equation}

Here, A$^{\pm} = (d_{\perp \theta} \pm 1)/2 $, $\theta_{F}$ = 35$^{\circ}$, $\theta_{A}$ = $\theta$ for $\theta \leq 90^{\circ}$ but  $\theta_{A}$ = 180$^{\circ}$ - $\theta$ with $\theta \geq$  90$^{\circ}$ and $d_{\perp \theta}$ = 6.0 to 3.0 (in this study). 
This means that $K_{\perp \theta}$ may have a different latitudinal dependence than the other diffusion coefficients, and can be enhanced towards the heliospheric poles by a factor $d_{\perp \theta}$ with respect to the value of $K_{\parallel}$ in the equatorial region of the heliosphere. We assumed $d_{\perp \theta}$ = 6 for the period before the solar polarity reversal, then gradually decreases it to 3.0. For a detailed theoretical discussion of this aspect of diffusion theory, see the extensive review by Schalchi (2009). For a motivation and application of this particular modeling approach, see Potgieter (2000), Ferreira et al. (2003a), Ngobeni \& Potgieter (2012), Nndanganeni \& Potgieter (2016), Potgieter et al. (2015) and Aslam et al. (2019a).

\begin{figure}[!htp]
\plotone{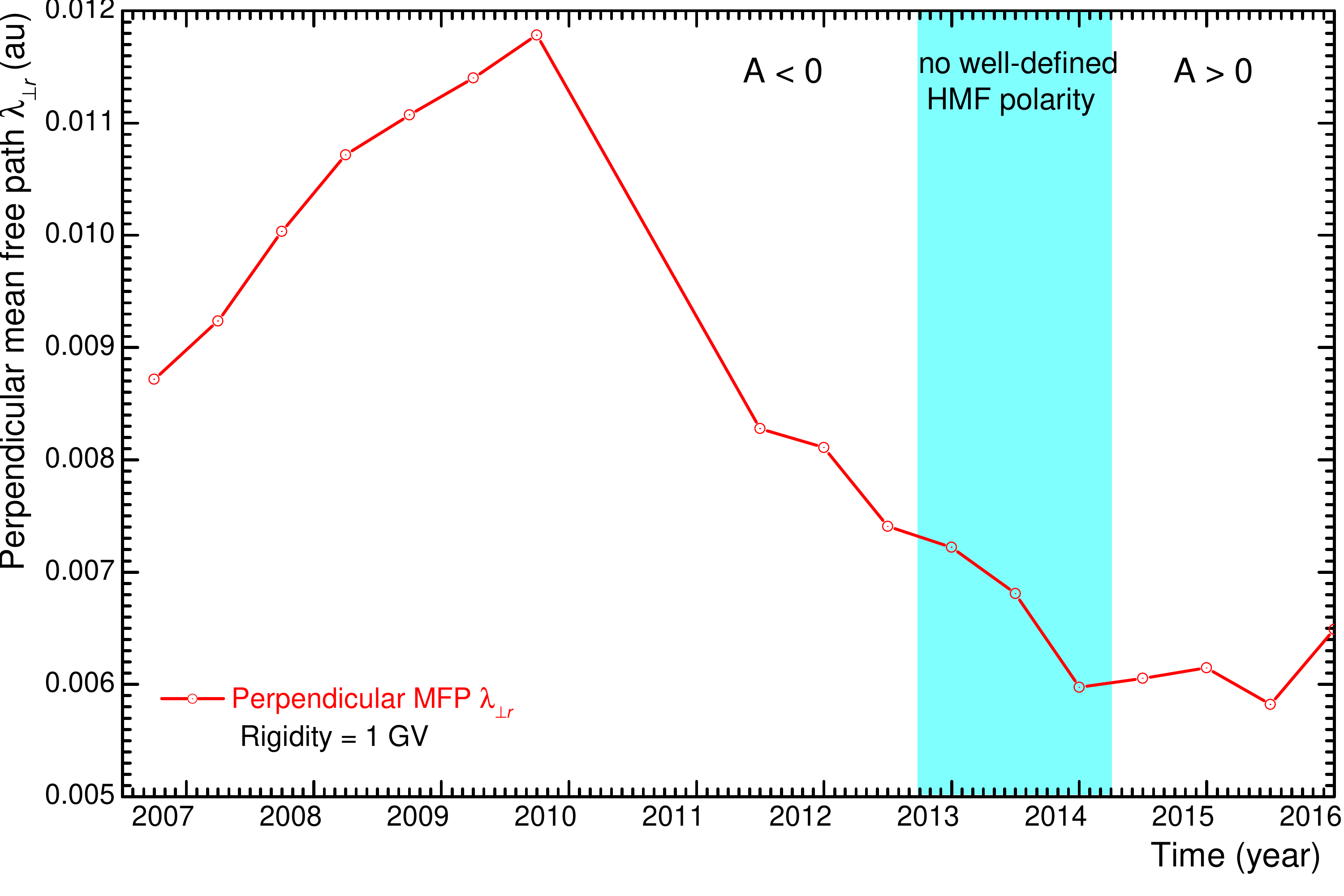}
\caption{Similar to Figure 6 but now for $\lambda_{\perp r}$. \label{fig7}} 
\end{figure}

Similar to the $\lambda_{\parallel}$, the calculated $\lambda_{\perp r}$ from mid-2006 to end of 2015 for 1.0 GV electrons and positrons is shown in Figure \ref{fig7}. This is not repeated for $\lambda_{\perp \theta}$.

It follows from these two figures that the MFPs show a gradual and systematic increase from mid-2006 to the end of 2009. See Aslam et al. (2019a) for how the other modulation parameter values were changed for the 2006 to 2009 period. 
As for the drift scale, the MFPs for January 2010 to April 2011 were obtained through extrapolation. The MFPs since 2011 are decreasing and reach the lowest level by the end of 2013, maintaining this low level even after the polarity reversal up to the end of 2015; see Aslam et al. (2020) for other modulation parameters and values of diffusion coefficients used for the period from May 2011 to December 2015.   
  
\begin{figure}[!htp]
\plotone{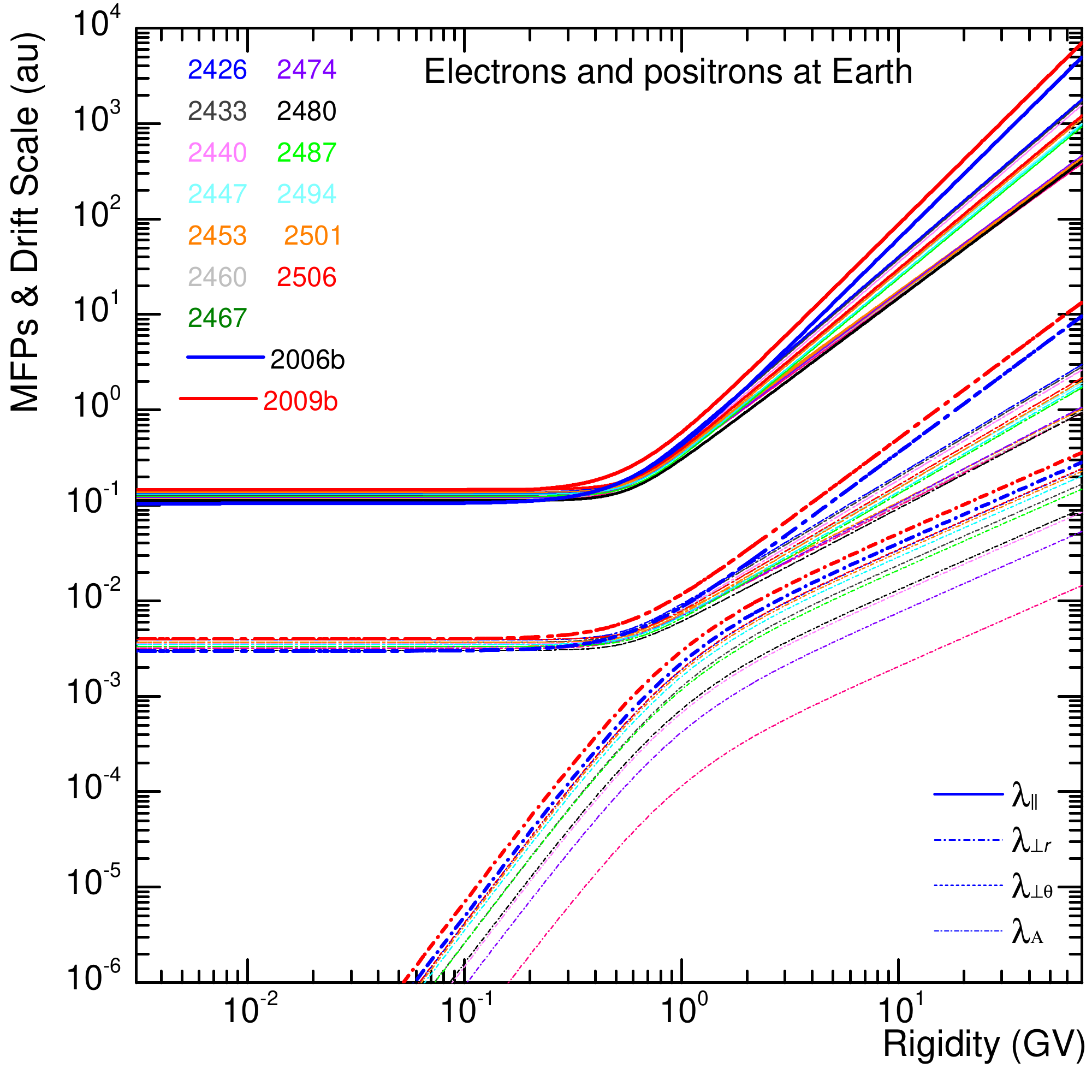}
\caption{ Rigidity dependence of the three MFPs and the drift scale at the Earth for electrons and positrons and how they varied from 2006 to 2015 indicated by the changing colors. The rigidity dependence of $\lambda_{\perp r}$ and $\lambda_{\perp \theta}$ are identical. These values follow directly from Equations (\ref {Eq4}), (\ref {Eq5}), (\ref {Eq6}). and (\ref {Eq7}).
\label{fig8}}
\end{figure}

In the next figure we focus on the computed rigidity dependence of the three diffusion coefficients and the drift coefficient at the Earth for the time periods used in this study. Again we use the corresponding MFPs and drift scale, all in units of au, as shown in Figure \ref{fig8}. 
Note that in Equations \ref{Eq5} and \ref{Eq6}, $c_{2\parallel}$ and $c_{2\perp}$ together with $c_{1}$ determine the slope of the rigidity dependence respectively above and below a rigidity $P_{k}$. For electrons and positrons, $c_{1}$ = 0, so that the slope of the rigidity dependence changes only above $P_{k}$. The values for $c_{2\parallel}$ = 2.25 and $c_{2\perp}$ = 1.688 remained unchanged for the solar minimum period of July 2006 to December 2009 (for the $\it{PAMELA}$ observations), but for the later periods from May 2011 to December 2015, we had to change both $c_{2\parallel}$ (from 1.98 to 1.75 for the 2013 and back to 1.88 for the December 2015), and $c_{2\perp}$ (from 1.287 to 1.103 for 2013 and back to 1.147 for the December 2015) in order to reproduce the $\it{AMS}02$ electron and positron observations. A full set of parameter values utilized to reproduce the $\it{PAMELA}$ spectra for mid-2006 to end-2009 is provided by Aslam et al. (2019a) and the parameter values used to reproduce Bartel rotation averaged for $\it{AMS}02$ electron and positron spectra from May 2011 - May 2017 is tabulated by Aslam et al. (2020), and not repeated here. 

According to the results shown in Figure \ref{fig8}, the $\it{AMS}02$ observations (2011 to 2015) require reduced rigidity dependent slopes for the three MFPs above about 1.0 GV compared to what was required for 2006 and 2009. This may seem inconsistent but keep in mind that a new solar activity cycles had started in 2010 and that the subsequent increasing turbulence could have developed differently than the declining values before minimum modulation (e.g. Zhao et al. 2018). This suggested behavior is consistent to how differently the tilt angles and HMF magnitude had evolved before and after solar minimum modulation as shown in Figure \ref{fig3}. 

\section{Discussion} \label{sec5}

The tilt angle, $\alpha$, of the HCS and the magnitude $B$ of the observed HMF at Earth are considered to be good proxies for solar activity in the numerical modeling of GCR modulation (Potgieter et al. 2014; Aslam \& Badruddin 2015). Their moving average values as depicted in Figure \ref{fig3}, and how they change in phase with solar activity, were utilized as essential time-varying modulation parameters and used as input for our 3D steady-state model in an attempt to make it as realistically as possible. The time dependence obtained from $B$ is incorporated into the three diffusion coefficients, and in the drift coefficient, all of which are assumed to scale proportional to $1/B$; the effect of this was shown in Figure \ref{fig5} from 2006 to 2009, when nothing else was changed with time. Of course, particle drifts are also responding to the value of $\alpha$, more so when GCRs drift inwards to the Earth along the HCS mostly through the equatorial regions of the heliosphere than when they drift inward through the polar regions of the heliosphere. In the polar regions the HCS plays a role when the tilt angles become very large typically when solar maximum activity is approached (see illustrations by Raath et al. 2015, 2016). However, in order to reproduce the observed $\it{PAMELA}$ and $\it{AMS}02$ spectra after 2009, we had to change the diffusion coefficients as well as the drift coefficient in addition to what time dependence was introduced through the moving averages of $\alpha$ and $B$ for each period. 
 
As shown in Figures \ref{fig6}, \ref{fig7} and \ref{fig8}, the diffusion coefficients in the model are required to change over time from 2006 to 2015. They show a fast increase from 2006 to 2009 (more than 30\% with respect to 2006) as compared to the drift coefficient, indicating that although the particle drifts played a meaningful role during this period, it was in fact diffusion-dominated, which made this solar minimum period different from previous solar minima. (See detailed discussions on this topic by Potgieter et al. 2014, Di Felice et al. 2017, Munini et al. 2018, and Aslam et al. 2019a). The diffusion coefficients show a fast and significant decrease after 2011 up to mid-2014 (starting of positive polarity), but remains at low levels until mid-2015, as an effect of high HMF values. Solar activity started to decrease by mid-2013, as is evident from the $\alpha$ time profile (Figure \ref{fig3}), but $B$ shows a gradual increase even after the complete polarity reversal. It is noted that the Hermanus NM in South Africa (cutoff rigidity of 4.6 GV) registered a minimum GCR intensity by the end of 2014 (see, http://natural-sciences.nwu.ac.za/neutron-monitor-data).   

We emphasize that in order to reproduce the observed electron and positron spectra, and the subsequent $e^{+}$/$e^{-}$ with time, especially during high solar activity, we had to scale the drift coefficient from a maximum level during the minimum activity phase to a minimum level during maximum solar activity (including the polarity reversal period), and then back to larger levels as solar activity decreased (Figure \ref{fig5}). This means that $\omega \tau$ in Equation \ref{Eq3} changes with time, following the solar activity cycle; in the model this drift reduction factor is determined by the value of the drift scaling $K_{A0}$ in Equation \ref{Eq4}. 
The electron and positron observations after 2011, as shown in Figure \ref{fig4}, thus require a continuous reduction from $K_{A0}$ = 0.90 down to $K_{A0}$ = 0.0 for the polarity reversal period. If this reduction is absent, that is, keeping the drift coefficient at high levels up to the reversal of the HMF polarity, a sudden and abrupt change in the ratio will be present in the model, that is, at the precise moment when the HMF polarity is reversed in the model. This behavior was found by Tomassetti et al. (2016) and reported previously by others when drift dominated models were explored and applied for the first time; see Webber et al. (1990), and Webber \& Potgieter (1989) for classic examples, and the review by Potgieter et al. (2001; see Figure 3). It has been known already since the $\it{Ulysses}$ mission (Heber \& Marsden 2001; Heber et al. 1999, 2002, 2003; Heber 2013) when electron to proton, and electron to helium ratios where observed over long periods that this kind of very sudden, abrupt changes was not observed so that drift models had been adjusted accordingly. In this context, see also Ferreira et al. (2003a,b), Ndiitwani et al. (2005); reviews by Ferreira \& Potgieter (2004), Heber \& Potgieter (2006). Potgieter (2014b) gave a short history of predicted and observed charge-sign dependent modulation. Further adjustments to drift modulation models came about when drift effects observed by $\it{PAMELA}$ have turned out to be less than predicted and observed for the $A<0$ polarity cycles before the solar minimum of 2009 (Di Felice et al. 2017; Potgieter 2017).

It is interesting that scaling the diffusion coefficients only with $B$, and the drift coefficient with $B$ and $\alpha$, with time seems overly simplified, at least for some phases of the solar activity cycle. As indicated above we had to make additional changes with time. A similar report was made by le Roux \& Potgieter (1995) who concluded that global merged interaction regions (GMIRS) forming beyond about 10 AU were required to reproduce the full amount of GCR modulation during 1977 to 1987. The effect of such a GMIR on $\it{AMS}02$ proton observations was reported and modeled by Luo et al. (2019). However, for the modulation period after 2009, it is not so obvious as after 1977 and 1987, how and when these GMIRS could have developed far beyond the Earth. Modeling such complicated time-dependent structures is beyond the scope of our model. In order to compensate for such effects on solar modulation, we may need to shorten and even change the averaging periods applied to $B$. It is also possible that the assumption that all diffusion coefficients scale spatially as $1/B$ is too rudimentary and more complicated functional forms are required e.g. as applied in models by Manuel et al. (2014) and recently by Caballero-Lopez et al. (2019), Moloto et al. (2018) and Moloto \& Engelbrecht (2020). 

Following up on Figure \ref{fig8}, it is of interest to note that the very flat rigidity slopes displayed below about 400 MV may be too simplified, requiring some adjustments but which can only be tested properly if electron and positron observations well below 100 MV are done simultaneously. In fact, basic turbulence theory (Bieber et al. 1994; Teufel \& Schlickeiser 2006) suggests that the rigidity dependence of the MFPs at these low values may be more complicated, even increasing below 1 MV but generally speaking always less than above 500 MV. The rigidity value where the prominent change in the slope occurs, usually accepted to be around 0.4 GV, may also be tested with additional observations. This situation is complicated by the fact that at these low kinetic energies, Jovian electrons become progressively dominant so that the actual GCR electron intensity at the Earth is obscured from viewing; see Vogt et al. (2018) and Nndanganeni \& Potgieter (2018). The relative modest changes with time in the diffusion MFPs below 500 MV, as displayed in Figure \ref{fig8}, is also somewhat surprising. As shown here, electron and positron modulation is dominated by diffusion below this rigidity (see also Potgieter 1996; Nndanganeni \& Potgieter 2016) so that it is reasonable to expect a larger time variations with changing solar activity. However, to investigate this in greater detail requires that $\it{PAMELA}$ electron spectra observed after 2009 to be published, whereas the published $\it{AMS02}$ spectra observed above 1.0 GV are not helpful in this context. Evidently, testing whether electron modulation is indeed completely dominated by diffusion at these low kinetic energies, and below what energy values this may happen (or in other words, down to what low energy value may particle drifts play a role in charge-sign dependent modulation), is not yet possible because of the absence of simultaneous and continuous measurements of GCR particles and their antiparticles in this energy range. Perhaps, the publication of $\it{PAMELA}$ electrons and positron spectra to the lowest observed kinetic energies until 2016 (Mikhailov et al. 2019), and relevant observations from balloon experiments (Mechbal et al. 2020) may assist in this regard.

The next phase of this study will be focused on the recent polarity reversal phase during solar activity maximum, by utilizing observations of GCR protons and anti-protons, helium nuclei and electrons and positrons but up to the present solar minimum phase, which as a very quiet $A>0$ cycle, should be quite interesting. Drift models predict that at kinetic energies below about 1 GeV/nuc, GCR protons, nuclei and positrons should be even higher than in 2009, but electrons and anti-protons should be lower, if modulation conditions were again to be as quiet as in 2006 to 2009; for such predictions, see Potgieter \& Vos (2017) and Aslam et al. (2019a,b). For preliminary modeling studies of anti-protons and the proton to anti-proton ratio, see Aslam et al. (2019c).

\section{Summary and Conclusions} \label{sec6}

The availability of both electron and positron spectra from July 2006 to December 2009 from $\it{PAMELA}$ together with $\it{AMS}02$ observations from May 2011 to December 2015, has made it possible to accomplish our primary objective, that is, to find ways of how to reproduce the observed positron to electron ratio over this time frame, including the reversal of the HMF polarity. This was done by applying a comprehensive and well established 3D numerical modulation model to compute full electron and positron spectra for this period. Doing so was preceded by establishing an electron VLIS based on applying the GALPROP code but adjusted to reproduce $\it{Voyager}$ 1 \& 2 electron observations around 10 MeV from beyond the HP together with $\it{PAMELA}$ and $\it{AMS}02$ observations at high enough energies where solar modulation can be considered negligible. For positrons no direct $\it{Voyager}$ observations exist so that the GALPROP based LIS was modified by Aslam et al. (2019a) to establish a VLIS. These VLIS were specified in the modulation model at the HP assumed to be at 122 au and then modulated using the physics as described in Parker's TPE (Equation \ref{Eq1}). This 3D modulation model, applied extensively before, includes particle drifts and a dynamic heliosheath (distanced between the TS and HP) related to the solar activity cycle. 

The moving averages of the HCS tilt angle, $\alpha$, and HMF magnitude, $B$, as observed at Earth, were used as proxies for time-varying solar activity. Changing the value of $B$ affects directly the diffusion and drift coefficients which scale proportional to $1/B$. However, the diffusion and drift coefficients had to be changed with time additionally in order to reproduce the individual spectra over time. The results displayed in Figure \ref{fig4} illustrate that the model can reproduce the $e^{+}$/$e^{-}$ over the KE range of 1.0 - 2.0 GeV as observed by $\it{PAMELA}$ and $\it{AMS}02$, from 2006 to 2015, including the HMF polarity reversal period. The exact same modulation parameters were used for both electrons and positrons in the model, with the only differences their corresponding VLIS and the change in the drift patterns that they follow during positive and negative polarity phases of the HMF.    

The consequent physics changes in the TPE were shown in Figures \ref{fig5} to \ref{fig8}, and is summarized as follows:
Both the diffusion and drift coefficients had to remain at high levels for the unusual solar minimum of 2006-2009 in order to reproduce the observed $\it{PAMELA}$ spectra of electrons and positrons. Even though drift played an important role in modulating these GCRs from 2006 - 2009, the period was essentially diffusion dominant. After 2009, we had to scale the diffusion and drift coefficients down systematically towards solar maximum until the HMF polarity reversal from A$<$0 to A$<$0 was complete. We had to keep the drift coefficient at its lowest level for the entire reversal phase from November 2012 - March 2014 (as the phase without a well-defined HMF polarity). 
From Figure \ref{fig8} follows how the rigidity dependence of the corresponding three MFP's and the drift scale changed with time, particularly the slope of the rigidity dependence above $P_{k}$, which is considered the rigidity where these modulation entities for electrons and positrons always change their rigidity dependence from a stronger to a weaker dependence, as rigidity decreases.
             
\acknowledgments The authors wish to thank the GALPROP developers and their funding bodies for access to and use of the GALPROPWebRun service. We acknowledge the use of HCS tilt data from Wilcox solar observatory's http://wso.stanford.edu website, and HMF data from NASA's OMNIWEB data interface http://omniweb.gsfc.nasa.gov. 
MDN acknowledges the SA National Research Foundation (NRF) for partial financial support under the Joint Science and Technology Research Collaboration between SA and Russia (Grant no:118915) and BAAP (Grant no: 120642). OPMA \& DB acknowledges the partial financial support from the post-doctoral program of the North-West University. RM acknowledges partial financial support from the INFN Grant \textquote{giovani}, project ASMDM. VVM acknowledges for RFBR and NRF research project No. 19-52-60003, and Ministry of Science and Higher Education of the Russian Federation under project No. 0723-2020-0040.

\end{document}